\documentclass[aps,preprint]{revtex4}%
\usepackage{amsfonts}
\usepackage{amsmath}
\usepackage{amssymb}
\usepackage{graphicx}%
\begin{document}
\title{Shock waves in virus fitness evolution }
\author{Fernando Goldenstein Carvalhaes and Carla Goldman}
\affiliation{Instituto de F\'{\i}sica, Universidade de S\~{a}o Paulo, CP 66318, 05315-970,
S\~{a}o Paulo, Brazil.}

\begin{abstract}
We consider a nonlinear partial differential equation of conservation type to
describe the dynamics of vesicular stomatitis virus observed in aliquots of
fixed particle number taken from an evolving clone at periodic intervals of
time \cite{novella 95}. The changes in time behavior of fitness function
noticed in experimental data are related to a crossover exhibited by the
solutions to this equation in the transient regime for pulse-like initial
conditions. As a consequence, the average replication rate of the population
is predicted to reach a plateau as a power $t^{-\frac{1}{2}}.$


\end{abstract}
\startpage{1}
\endpage{9}
\maketitle


The evolution of RNA viruses has been extensively studied in several systems,
including vesicular stomatitis virus (VSV), foot-and-mouth disease virus
(FMDV) and phage $\phi-6$ \cite{domingo}, \cite{novella 03},
\cite{elena/lenski}. This study is facilitated by the fact that virus clones
present very high replication and mutation rates so a highly heterogeneous
population - a \textit{quasispecies}, as referred in virology
\cite{eigenquasispecies} - can be derived from a clone in relative short
periods of time. Some of the experiments done with VSV populations were
directed to study how the process of transmission of fixed amounts of
pathogens from host to host (much like as in a flu, for example) is expected
to influence the process of evolution of \ the \textit{fitness} of a clone.
\cite{novella 95},\cite{novella 99},\cite{novella 04}.

In order to investigate in a theoretical framework some aspects of the
interplay between these two processes we model the system dynamics, in analogy
to the works in Refs. \cite{tsimring} and \cite{aafif}, by a Markovian process
that take place in a properly defined one-dimensional (phenotypic) ordered
space. The research interest is both conceptual and motivated not only by
questions in virology but also by their application in different fields, as
traffic flow, intracellular transport, molecular motors, and others that can
be connected to the motion of self-driven many-particle systems \cite{review}.
We restrict our investigation, however, to systems of conservative nature.

The experimental procedure used in studies of VSV follows the original
protocol proposed and described in details in Ref. \cite{holland 91}. In
short, at instant $n\tau$ of the $n-th$ transmission, $n=1,2,3,...,$ a number
$N_{n}(0)\sim10^{5}$ of viral particles is taken from an evolving clone (ec)
of VSV of initial low fitness and inoculated into a recipient containing a
known amount of non-infected cells. The viruses are allowed to replicate for a
period $\tau=$ $24$ hours, reaching a total of $N_{n}(\tau)$ particles at the
end. This accomplishes for one\textrm{\ }\textit{evolutive passage} along
which selection and mutation processes are expected to take place\textit{.}
Then an aliquot with $N_{n+1}(0)\sim10^{5}$ particles is taken from the ec and
the procedure is repeated using new (non-infected) cells for the next passage.
The \textit{fitness} $F_{n}(\tau)$ of the ec at the end of the evolutive
passage $n$ can be expressed as
\begin{equation}
F_{n}(\tau)\equiv\frac{N_{n}(\tau)}{N_{n}(0)}=\exp\left\{  \int_{n\tau
}^{(n+1)\tau}r_{e}(t)dt\right\}  =\exp\left\{  r_{e}(t_{n})\tau\right\}
\label{Fn(tau)}%
\end{equation}
where the growth rate $r_{e}(t)$ of \ the (ec) is assumed a continuous
function of time in the considered interval so that the second equality
follows from the mean value theorem of calculus for $t_{n}$ within the
interval $n\tau<t_{n}$ $<(n+1)\tau$. By taking the initial quantity in the
aliquot $N_{n}(0)=N(0)\equiv N$ independent of $n,$ one expects to mimic in
this way the process of infection by transmission of fixed amounts of
pathogens.\textit{\ }The qualitative properties of the population of these $N$
individuals at each passage, and in particular the value of $r_{e}(t_{n})$,
are expected to change considerably from one passage to another. The aim of
the experimental procedure is to get information on the time behavior of the
relative quantity $r(t_{n})=r_{e}(t_{n})-r_{w}$ measured with respect to the
growth rate $r_{w}$ of a non-evolving \textit{wild type}\textrm{\ }(wt)
population taken as a standard.

Two sets of data obtained for $r(t_{n})$, measured in units of $\tau^{-1}$
\cite{fitness vs rate} for $n=1,2,...$\ , are reproduced in \emph{Fig.1
}\cite{novella 95}. One of these (stars) shows a fast increasing regime at
initial values of $n$ followed by a region where $r(t_{n})$ grows more slowly
as $n$ increases. The other set (squares) apparently follows just the fast
regime at initial times followed by a region where data present strong
fluctuations. A plateau (not observed in these data) is expected to be
attained around passage $80$, as reported latter in Ref. \cite{novella 99}.
Yet, despite the fact that the data presents immense fluctuations at large $n$
so the existence \ (and localization) of a plateau could still be debated, one
should not expect that the rate of population increasing grows indefinitely
due to metabolic constraints and limitations of resources \cite{aafif}.%
\[
Figure1
\]
\qquad\qquad

A few theoretical works have already been devoted to describe the observed
dynamics of VSV as in Refs. \cite{tsimring},\cite{aafif},\cite{fisher
discreto},\cite{rouzine} among others. Although the description in
\cite{tsimring} does not allow for a finite asymptotic limit, it predicts
correctly the two-phase regimes, as observed in experiments, but conditioned
to the presence of random distributions at initial times. The expected limit
is obtained in Ref. \cite{aafif} by a modified version of the model in
\cite{tsimring}. Our motivations to study these data once again from a
theoretical perspective came essentially from the possibility to investigate a
few questions concerning the interplay between the dynamics of fitness
evolution and that imposed by the particular experimental environment; namely
(1) what would be an appropriate way to account for effects of adaptability of
individual components to the particular transmission mode; (2) what are the
possible causes and in what conditions one should expect two-phase regimes;
and (3) how the kinetics of fitness is affected by the form of initial
population distribution.

The key point that have guided our formulation is the fact that $r(t_{n}%
)$\ refers to a property of the population in the aliquot chosen at the
beginning of each passage $n.$ It seems appropriate then to consider $\tau$ as
the unit of time and in connection to this, a process that describes fitness
evolution but conserves the total number $N$ of individuals.

To this end, the particles are considered distributed into distinct components
(or subpopulations) indexed by $k=0,1,2,...$each characterized by a fixed
fitness value. Let $N_{n,k}$ be the number of particles of the $k$-component
present in the aliquot at\ time $n\tau$. \emph{ }The dynamics of $N_{n,k}$
from one passage to another is mapped onto a jumping process that take place
in an appropriate fitness space where to each position $k$ we assign a value
$k\lambda$ with $\lambda$ a positive constant. Here, $k\lambda$ is defined as
the logarithm of fitness of the $k$-component. Accordingly, $k\lambda$ may
depend on its intrinsic growth rate $r_{k}$ and on the (fixed) time interval
used to attribute a fitness measure to it. Such dependence, however, does not
need to be elucidated here. Within this view $N_{n,k}$ is seen as the
occupation number at site $k,$ at time $n\tau$. The corresponding
concentration $C_{n,k}$\ is given by $C_{n,k}=N_{n,k}/N$ with $\sum_{k}%
N_{n,k}=N.$ Jumping of particles that can give rise to changes in occupation
numbers are assumed to occur between neighboring sites at any instant within
the interval $\tau$ between any two passages. Moreover, particles at each site
are supposed to behave independently from each other; thus $N_{n,k}$ satisfies
the following recurrence relation%
\begin{equation}
N_{n+1,k}=N_{n,k}+p_{n,k-1}N_{n,k-1}+q_{n,k+1}N_{n,k+1}-p_{n,k}N_{n,k}%
-q_{n,k}N_{n,k}\label{markov}%
\end{equation}
for $n=1,2,...$. The quantities $p_{n,k}$ and $q_{n,k}$ are the transition
probabilities between neighboring sites. $p_{n,k}$ is the probability for
increasing the fitness of the whole population by increasing occupation at
position $(k+1).$ $q_{n,k}$ is the probability for decreasing the fitness of
the whole population by increasing occupation at position $(k-1).$ The model
is properly defined when we specify the forms of $p_{n,k}$ and $q_{n,k}$. We
choose%
\begin{equation}
p_{n,k}=P-\alpha C_{n,k}\label{p}%
\end{equation}%
\begin{equation}
q_{n,k}=Q+\alpha C_{n,k}\label{q}%
\end{equation}
Parameters $P,$ $Q$ and $\alpha$ are positive numbers such that $0\leq
P\leq1,$ $Q=1-P$ and $\alpha\leq P.$ This choice was inspired by the dynamics
of clannish random walkers, proposed by Montroll to describe a related process
of gas separation \cite{montroll}. In population dynamics, it resembles
extended logistic models already considered in the literature \cite{fisher
discreto} \cite{saldana 03}, \cite{elm} to study effects of competition among
the diverse subpopulations.

Our reasoning is connected to the first question posed above. It comes from
observation that at each transmission, some of the components of high growth
rates may happen to be less adapted to the experimental environment so to
appear depleted at the time of sampling if compared to components of
low\textrm{\ }growth rates. This point, which has already been discussed in
the literature \cite{holland 90} for considering coinfections and
superinfections processes \cite{fisher discreto}, \cite{saldana 03},
emphasizes the necessity to account for the interplay between the intrinsic
dynamics of virus and that imposed by the particular transmission mode
\cite{saldana 03}. Therefore, in our formulation we use the concentrations of
the components present in the aliquot at each time (or at the time of each
passage) as the proper measure of their relative adaptability at that time.
The strength with which this affects the dynamics of fitness is regulated by
parameter $\alpha.$ At the genotypic level (microscopic) this parallels
effects of random genetic drift by sampling in populations of limited size
\cite{kimura}. Parameter $P$ is a\ measure of the fraction of the progeny of
component with fitness $(k-1)\lambda$\ present at passage $n$\ that in the
absence of the terms in $\alpha,$ would appear mutated into component of
fitness $k\lambda$ in the following passage.\ We attribute a numerical value
to $P$ as characteristic of each clone of VSV, at each state of dilution
regarding the multiplicity of (cell) infection (m.o.i.).

We take $\lambda\rightarrow0,$\ $\tau\rightarrow0$\ keeping the ratio
$\dfrac{\lambda}{\tau}\rightarrow$\ $\Gamma$\ finite. In these limits,
$C_{n,k}$\ becomes a function $C(x,t)$\ of two continuum variables,\ the
position $k\lambda\rightarrow x$\ and time $n\tau\rightarrow$\ $t$\ for $k$
and $n\rightarrow\infty$ and Eq. (\ref{markov}) converges to a quasi-linear
PDE of conservation type,%

\begin{equation}
\frac{\partial C}{\partial t}+(a-bC)\frac{\partial C}{\partial x}=0
\label{eqcont}%
\end{equation}
where $a\equiv\Gamma(2P-1)$ and $b\equiv4\alpha\Gamma$ are measured in units
of $t^{-1}$. In Eq. (\ref{eqcont}) number conservation appears as a
\textit{local} property \cite{foot2}.

The connection of the model with the measurements of the average fitness
performed in the experiments is made by expressing the exponent in
Eq.(\ref{Fn(tau)}), up to a constant $r_{w},$ as
\begin{equation}
r(t_{n})\tau\equiv\left\langle x\right\rangle (n\tau+\tau)-\left\langle
x\right\rangle (n\tau) \label{connection}%
\end{equation}
where in the above limits $\left\langle x\right\rangle $ converges to a
continuous function of time. Our proposal consists in approaching this by the
average
\begin{equation}
\left\langle x\right\rangle (t)=%
{\displaystyle\int}
xC(x,t)dx \label{xmedio}%
\end{equation}
and, consequently
\begin{equation}
r(t)=\frac{d}{dt}\left\langle x\right\rangle (t) \label{r(t)}%
\end{equation}
which is the quantity of interest. The solutions to Eq. (\ref{eqcont}) for
certain initial conditions $C(x,0)$ can be examined using the method of
characteristics \cite{haberman}, \cite{evans}. We first consider a rectangular
pulse such that $C(x,0)=0$ for $x_{0}\leq\Lambda;$ $C(x,0)=h$ for
$\Lambda<x\leq\Lambda+m$ and $C(x,0)=0$ for \ $x>\Lambda+m.$ The quantities
$h,m$ and $\Lambda$ are constants. $m$ is the initial dispersion and $\Lambda$
controls the average$\left\langle x\right\rangle (0)$ at initial time.
Normalization of $C(x,t)$ at all times is ensured by setting $h\cdot m=1.$ For
this choice of initial conditions, there is occurrence of two shocks of
characteristics, one at time $t_{s(1)}=0$, at $x_{s(1)}(0)=\Lambda,$ and
another at time
\begin{equation}
t_{s(2)}=\frac{2m}{hb}=\frac{2m^{2}}{b} \label{t2shock}%
\end{equation}
at position $x_{s(2)}(0)=m(\dfrac{2a}{bh}-1)+\Lambda.$ This second shock
happens when the first shock front encounters the family of characteristics
$x_{r}(t)=m+\left[  a-bC\right]  t+\Lambda$ in the region of rarefaction that
emerges from $x_{0}=\Lambda+m.$ The curves $x_{s(1)}(t)=\Lambda+\left[
a-b/2m\right]  t$ and $x_{s(2)}(t)=at-\sqrt{2bt}+\Lambda+m$ for $t>t_{s(2)}$
that describe the dynamics of the two shock fronts are determined from direct
integration of velocities $\overset{\cdot}{x}_{s(i)},$ $i=1,2,$ prescribed by
Rankine's condition \cite{evans}. The solutions $C(x,t)=(\Lambda+m-x+at)/bt$
at the rarefaction interpolate the regions where $C=0$ and $C=h.$ The solution
profile $C(x,t)$ for all $x$ is represented in \emph{Fig.2} at different
instants of time. Observe that because $C(x,t)$ has distinct dependence on
$t$, before and after the second shock, the corresponding
averages$\left\langle x\right\rangle (t)$ and consequently $r(t)$ computed in
these different intervals are expected to$\ $be distinct functions of time as
well so as to exhibit a crossover at $t=t_{2s}$. The mathematical reason for
this is, of course, the presence of nonlinear terms in $\alpha$ in the
evolution equation (\ref{eqcont}). In fact, before shock, we obtain%

\begin{equation}
r(t)\equiv r_{1}(t)=a-\frac{b}{2m}+\frac{b^{2}}{12m^{3}}t \label{r1ret}%
\end{equation}
reproducing the observed linear behavior of the growth rate of clones of VSV
at small times. The novel feature comes out from the solutions after shock
when $r(t)$ approaches a constant equal to $a$ as a power $t^{-1/2}$ since for
$t\geq t_{2s}$%

\begin{equation}
r(t)\equiv r_{2}(t)=a-\frac{\sqrt{2b}}{3}\frac{1}{\sqrt{t}} \label{r2ret}%
\end{equation}
We conjecture that such a crossover at the transient regime is in the origin
of changes in time behavior observed in the evolution of clones of VSV. Notice
that for another initial pulse-like (normalized) distribution with a
triangular profile, such that $C(x,0)=\dfrac{hx}{m}$ for $0<x\leq m$ and
$C(x,0)=0$ in all other regions, the shock of characteristics occurs at
$t_{s}=\dfrac{m^{2}}{2b}$ \ and the solutions to $r(t)$ show the same behavior
$t^{-1/2}$ in approaching the constant $a$ at large times. Numerical solutions
obtained by iterating Eq. (\ref{markov}) for Gaussian initial distributions
display similar behavior (results not shown).

Qualitative features exhibited by $C(x,t)$ in \emph{Fig.2 }can be regarded as
an evidence of a compromise between a tendency for maintenance of the
structure of the initial distribution, that determines the dynamics within the
first stage (before shock)\ and that of spreading into many components of
relative low concentrations at latter times (after shock). Thus, we understand
that these results support arguments in the literature according to which the
first stage of\ VSV fitness increasing is controlled by \ mechanisms of
natural selection acting on the evolving population as a whole whereas the
dynamics at the second stage is characterized by bottleneck effects acting on
low concentration components \cite{novella 99}. Expression (\ref{t2shock}) for
$t_{2s}$ suggests that observation of these two regimes, however, is
conditioned either to the strength of this compromise measured by $\alpha
$\ (or, $b$) and to the extent of the initial dispersion $m.$ This is in
agreement with the fitting of data in \emph{Fig.1} into $r_{1}(t)$ for one of
the two sets (square), which present a high value for $t_{2s},$ and that into
$r_{2}(t)$ for the other (star) \cite{foot1} for which $t_{2s}$ results to be
relatively small. Thus, the first set appears to be localized in the region of
linear growth that precedes the shock. The other set (star) seems to reproduce
the behavior after shock. Different results were discussed in Ref.
\cite{tsimring}; according to these a condition for observation of two-phase
regimes would be the presence of random distributions at initial times (not
pulse like).

In conclusion, the model proposed here to describe the interplay between the
process of transmission and that related to intrinsic fitness evolution allows
one to interpret the observations in colonies of VSV in terms of a crossover
between two time regimes and elucidate the conditions for such observation. In
addition, the model predicts that the asymptotic limit of $r(t)$ is approached
as a power $t^{-1/2}$. The high quality of the fitting with $r_{2}(t)$ in
\emph{Fig.1} strongly supports this prediction. This must be contrasted,
however, with the results in Ref. \cite{aafif} according to which the temporal
behavior of $r(t)$ appears to be determined by a combination of two
exponential functions. Hence, a more accurate study of this region with
emphasis on an interplay between theoretical and experimental views shall be
crucial to decide on these matters. In turn, this might be relevant to studies
in virology for predicting novel aspects of the course of evolution of viral populations.

\vspace{0.5in}

\textbf{Acknowledgments}

We are very grateful to Dr. Isabel Novella for helping us to understand some
details of the experimental procedure and for her comments and suggestions to
the original manuscript. We also deeply acknowledge our colleagues Alexei M.
Veneziani, Gustavo B. de Oliveira and Domingos H.U. Marchetti for introducing
us into the subject of conservation laws. FGC acknowledges the financial
support by Conselho Nacional de Desenvolvimento Cient\'{\i}fico e
Tecnol\'{o}gico (CNPq), Brazil.

\emph{\bigskip}\textbf{Figure Caption}

1 - Experimental data reproduced from Fig. 1A MARM clone C (square) and Fig.
2B MARM clone D (star) of Ref. \cite{novella 95} . The lines represent the
best fittings of data using the functions in Eq.(\ref{r1ret}) (dashed) and
(\ref{r2ret}) continuum \cite{foot1}. \textit{Inset}: the best fitting with
$r_{2}(t)$ for clone D in approaching the asymptotic regime (few data points
were excluded in this fitting as an the attempt to minimize fluctuations and
capture the polynomial behavior at the relevant region).

2 - Solution profile for initial rectangular pulse shown at different instants
of time.\ (a) At $t=0$; (b) at an instant before the second shock
$(0<t<t_{2s})$: $C(x,t)=n$, for $x_{1}(t)<x<x_{2}(t)$, rarefaction in the
region $x_{2}(t)<x<x_{3}(t)$; (c) at an instant after the second shock
$(t>t_{2s})$. Here $x_{1}(t)=\Delta+(a-\dfrac{b}{2m})t$, $x_{2}(t)=\Delta
+m+(a-\dfrac{b}{m})t$ and $x_{3}(t)=\Delta+m+at$.


\begin{thebibliography}{99}
\bibitem {fitness vs rate}If $\tau$ equals to 1 day, as usual, then the rate
$r(t_{n})$ in units of (days)$^{-1}$ at discrete times $t_{n}$ is obtained
directly from the logarithm of the fitness. This is precisely the quantity
reported in experimental works.
\bibitem {foot2}In Refs. \cite{tsimring} and \cite{aafif}, number conservation
is ensured as a global property by the introduction of an additional parameter
into the evolution equation.
\bibitem {foot1}Because of data fluctuation, we established a compatibility
criteria in these fittings for combining the use of the two functions
$r_{1}(t)$ and $r_{2}(t)$. This imposes a compromise between the estimates for
parameter $b$ obtained from these fittings and the resulting values for
$t_{2}$ using (\ref{t2shock}) for an arbitrary value of $m$. For one of the
sets (square) the best fitting was achieved with $r_{1}(t)$ excluding the last
two points to avoid the region of high fluctuations in data. For the other
(star) the best fitting was achieved with $r_{2}(t),$excluding the first point
to avoid divergence at $t=0$\ and the last point in the region of fluctuation.
\end{thebibliography}
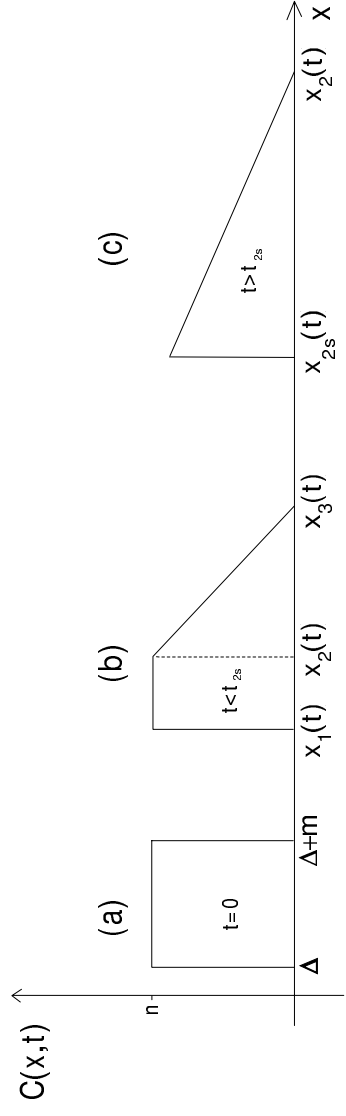
\end{document}